%% file: template.tex
\documentclass{vgtc}                          

\graphicspath{{figures/}} 

\usepackage{times}                     

\usepackage{tabu}                      
\usepackage{booktabs}                  
\usepackage{graphicx}
\usepackage{subcaption}
\usepackage[export]{adjustbox}

\usepackage{mathptmx}                  
\usepackage{amsmath}

\usepackage{tikz}
\usetikzlibrary{shapes.geometric, arrows, positioning, calc, fit, backgrounds}

\onlineid{0}

\vgtccategory{Research}

\vgtcinsertpkg

\title{Textured Word-As-Image illustration}

\author{Mohammad Javadian Farzaneh\thanks{e-mail: mohammad.javadian@sabanciuniv.edu}\\ %
        \scriptsize Sabanci University, Istanbul, Turkey %
\and Selim Balcisoy\thanks{e-mail: balcisoy@sabanciuniv.edu}\\ %
     \scriptsize Sabanci University, Istanbul, Turkey %
}

\abstract{
    In this paper, we propose a novel fully automatic pipeline to generate text images that are legible and strongly aligned to the desired semantic concept taken from the users’ inputs. In our method, users are able to put three inputs into the system, including a semantic concept, a word, and a letter. The semantic concept will be used to change the shape of the input letter and generate the texture based on the pre-defined prompt using stable diffusion models. Our pipeline maps the texture on a text image in a way that preserves the readability of the whole output while preserving legibility. The system also provides real-time adjustments for the user to change the scale of the texture and apply it to the text image. User evaluations demonstrate that our method effectively represents semantic meaning without compromising legibility, making it a robust and innovative tool for graphic design, logo creation, and artistic typography.

} 


\begin{document}


\firstsection{Introduction}

\maketitle

Artistic text generation includes text stylization and semantic typography. Text stylization focuses on the generation of text-effect overlays that appear on the texts such as shadows, outlines, colors, and textures. On the other hand, semantic typography is the art of using typography to visually reflect the meaning of a word by changing the size, shape, and other typographic elements of the letters\cite{iluz2023word}. These changes will align the visual representation of the text more with the desired concepts \cite{bai2024intelligent}.
Several techniques in textual visual design change the representation of the word in a way that conveys the meaning of the text. Semantic typography, calligrams, and ornamental typefaces are some examples of these methods. These methods have various applications in the generation of logos, greeting cards, and sign designs because visually expressing words has powerful rational and aesthetic applications.

In the past, humans performed text stylization manually. However, with advances in deep learning models and generative AI, researchers are now working to automate the process of calligram generation and semantic typography. For example, Zou et al. \cite{zou2016legible} proposed an automatic method for generating calligrams by extracting the required shape from an image and modifying the shape of each letter to better fit the desired form. Another study in semantic typography is the work of Iluz et al. \cite{iluz2023word}, which introduced a method for altering the shape of a word’s letters based on a user-provided semantic concept or the word’s inherent meaning using generative models. The output of their work is an image of a word with modified letters that represent the intended semantic concept. However, the authors chose to keep the results in black and did not apply any texture to the text.

The goal of this paper is to propose a new pipeline that integrates and synthesizes the concepts of artistic text stylization and semantic typography. Artistic text rendering relies heavily on artistic image rendering techniques. Proposed algorithms in this field may start with stroke-based methods \cite{hertzmann2001paint, hertzmann1998painterly} and extend to deep learning models \cite{Gatys_2016_CVPR, isola2017image}. In the era of AI-generated content (AIGC), thanks to powerful large-scale models \cite{podell2023sdxl, rombach2022high, ho2020denoising}, we are able to generate highly appealing text styles and typographies.

The developed pipeline takes input from users in the form of a semantic concept, word, and letter. The semantic concept is used to generate a subtle texture and modify the shape of the letter to visually convey its meaning. The proposed pipeline can have various applications in fields such as graphic design. It modifies the shape of letters and applies textures to text images, making it suitable for creating artworks and logo designs, having significant impact on advertisements and multimedia content creation \cite{bai2024intelligent}. 

\begin{figure*}[ht]
	\begin{center}
			\input{figures/dataflowdiagram}
	\end{center}
	\caption{The Proposed pipeline takes three inputs including "Semantic Concept", "Word", and a "Letter". In the provided example the inputs are "TREE" for semantic concept and "NATURE" for word choice, and the user has chosen letter "T" as the input. Then these inputs are sent to the "Word-As-Image" module\cite{iluz2023word} to produce a text image with the stylized letter chosen by the user. Also, the "Semantic Concept" will be used to generate a subtle texture using a stable diffusion model\cite{podell2023sdxl}. For instance, the "Word-As-Image" module has produced a word "NATURE" and changed the letter "T" into tree. Also, the stable diffusion module has produced a pattern of trees. Then, the generated texture will be mapped to the text image using the developed "Texture Mapping" module to generate an initial output. After output generation, the user is able to use texture scaling to better fit the texture and the color picker to change the default color of the background.}
	\label{fig:dataflow_diagram}
\end{figure*}
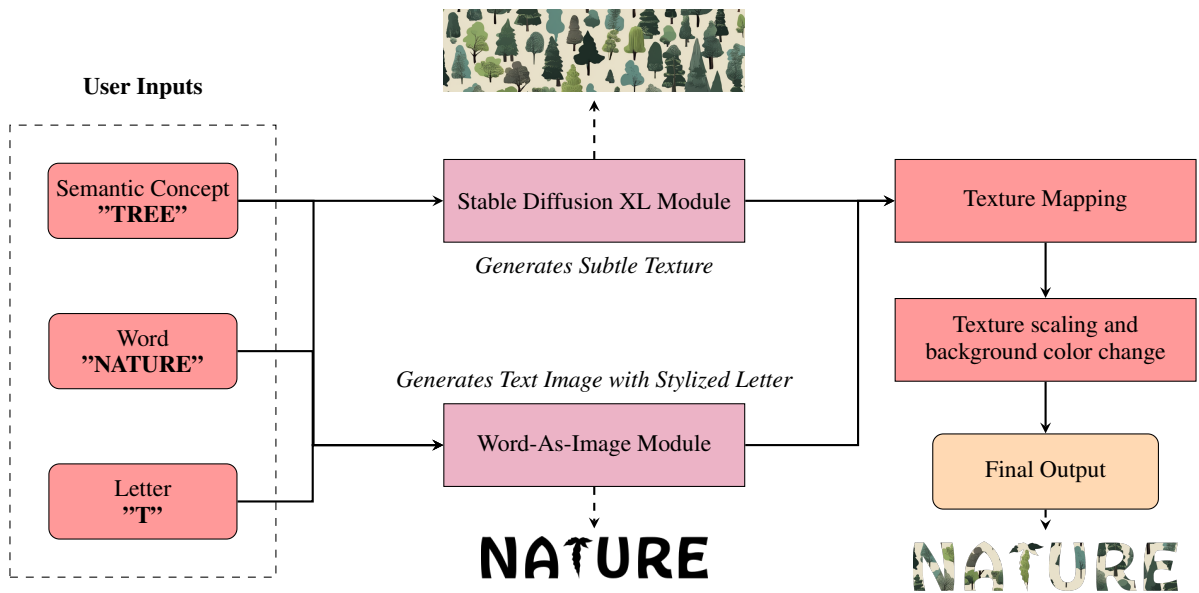
\section{Related Works}
With advancements in computational techniques, text stylization extends beyond font design, focusing instead on creating contextually adaptive visual representations. One of the methods that is widely used for artistic text generation is style transfer \cite{wang2022artistic}. Researchers have conducted several studies on style transfer, including the work of Yang et al. \cite{Yang_2017_CVPR, yang2018context, Yang_2019_ICCV} and Wang et al. \cite{wang2019typography}. For instance, in one of the works by Yang et al. \cite{yang2018context}, the authors proposed a two-step method to generate a style for plain text using any arbitrary style image without using deep-learning or generative models. In the first step, they apply the style of the image to the text, and in the second step, they determine the optimal position of the stylized text on the background image. To apply the style, a guidance map is first extracted from the style image. This map is a simplified version of the original image that retains important structural features such as edges and patterns. After extracting the style, the shape of the letters is modified to better integrate the style with the text.

In artistic typography, some methods try to embed images in letters or replace the letters with similar images. Research by Tendulkar et al. \cite{tendulkar2019trick} is one of them. In their method, the authors ask users to provide a text and a theme that will be integrated into the text. For instance, if the user enters the word "EXAM" with the theme of "Education," their method attempts to replace each letter with an icon or image that represents the theme while also matching the shape of the letters. The icons are sourced from a project named NEON and do not fall under the category of AI-generated content. Another method that works through blending images into the letters is the work of Zhang et al. \cite{zhang2017synthesizing}, creating artistic texts through blending images into the shape of the letters. Their process consists of three main steps. First, they perform segmentation on the letters, converting each letter into smaller parts called strokes. After segmentation, the system searches for similar images in both semantics and shape. In the image selection phase, these images are then combined to form the letters and, ultimately, the words. Berio et al. \cite{berio2022strokestyles} introduced StrokeStyles, a method for stylizing fonts. They are apply segmentation to extract the strokes and style them based on the user's preferences. They used Medial Axes Transformation (MAT) \cite{blum1967transformation}, which enables them to stylize each letter more effectively.

These days researchers are utilizing generative models to create visually appealing text styles \cite{peong2024typographic}. Tanveer et al. \cite{tanveer2023ds} and Mu et al. \cite{mu2025fontstudio} are some examples that have used generative models specially diffusion models to stylize the letters based on the semantic meanings. Mu et al. \cite{mu2025fontstudio} developed FontStudio, a generative model designed to create high-quality artistic fonts. It focuses on generating style effects for letters using a Shape-Adaptive Diffusion Model (SDM) based on user's desired letter and entered style prompt.

One of the projects that uses semantic concepts to create artistic text images is the work of Zhang et al. \cite{zhang2022creating}. They have proposed a method to automatically generate word paintings that integrate texts into the images in a way that represents semantic meanings and aesthetics. Their proposed pipeline gets an input image from the user to extract its features and generate keywords insert into the image. In another similar approach, Zou et al. \cite{zou2016legible} proposed a method to automatically generate calligrams based on provided images including three main steps: path generation, letter alignment, and letter deformation. In the next section, we will introduced the proposed pipeline.
\section{Methodology}
\subsection{Overview of the Pipeline}
We proposed a fully automatic pipeline capable of generating a text image with a stylized letter and subtle texture based on the user's desired semantic concept. First, the user enters three inputs into the system: a semantic concept, a word, and a letter from the word. The pipeline then modifies the shape of the chosen letter to represent the semantic concept using the "Word as Illustration" model \cite{iluz2023word}. Additionally, our method leverages the semantic concept to generate a subtle texture using the Stable Diffusion XL model \cite{podell2023sdxl}. After generating the subtle texture and the raw text image with the stylized letter, our method maps the texture onto the text image to produce the final output. Moreover, if the user is unsatisfied with the result, they can scale down the texture and change the background color to create a more visually appealing image. An overview of the proposed pipeline is shown in \cref{fig:dataflow_diagram}.

We developed a user interface with three text boxes that accept three inputs from the user. The first input is the semantic concept, which is used both to generate the subtle texture and to modify the shape of the letter in a way that visually represents the concept. The second input allows the user to enter their desired word, while the third input specifies the letter to be stylized. When the user clicks the "Generate" button, an image containing text with a stylized letter and texture is generated. After image generation, the user can click the "Adjust Image" button, changing the default white background or scaling of the texture, or download the image in PNG format.
\subsection{Letter Reshaping}
After the user fills in the three inputs—semantic concept, word, and letter—they click the "Generate" button. Upon clicking this button, the system calls the first part of the pipeline, which generates a text image containing the shaped letter. For this part of the pipeline, we used the "Word as Illustration" model \cite{iluz2023word}. Their approach first converts each letter into a vector shape representation using the FreeType font library. They extract the outlines of the letters and convert them into a uniform cubic Bézier curve representation to ensure consistency across different fonts. This method adds control points to the letter outlines. Since some letter outlines may have an inadequate number of control points, they apply a step called Control Point Subdivision to segment each Bézier curve based on the arc length of the segments.

\begin{figure*}[ht]
  \centering
  \begin{subfigure}[b]{\textwidth}
    \centering
    \includegraphics[valign=c,width=0.30\textwidth]{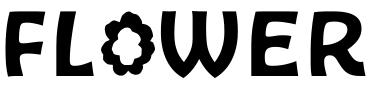}\hfill
    {\Large$+$}\hfill
    \includegraphics[valign=c,width=0.15\textwidth]{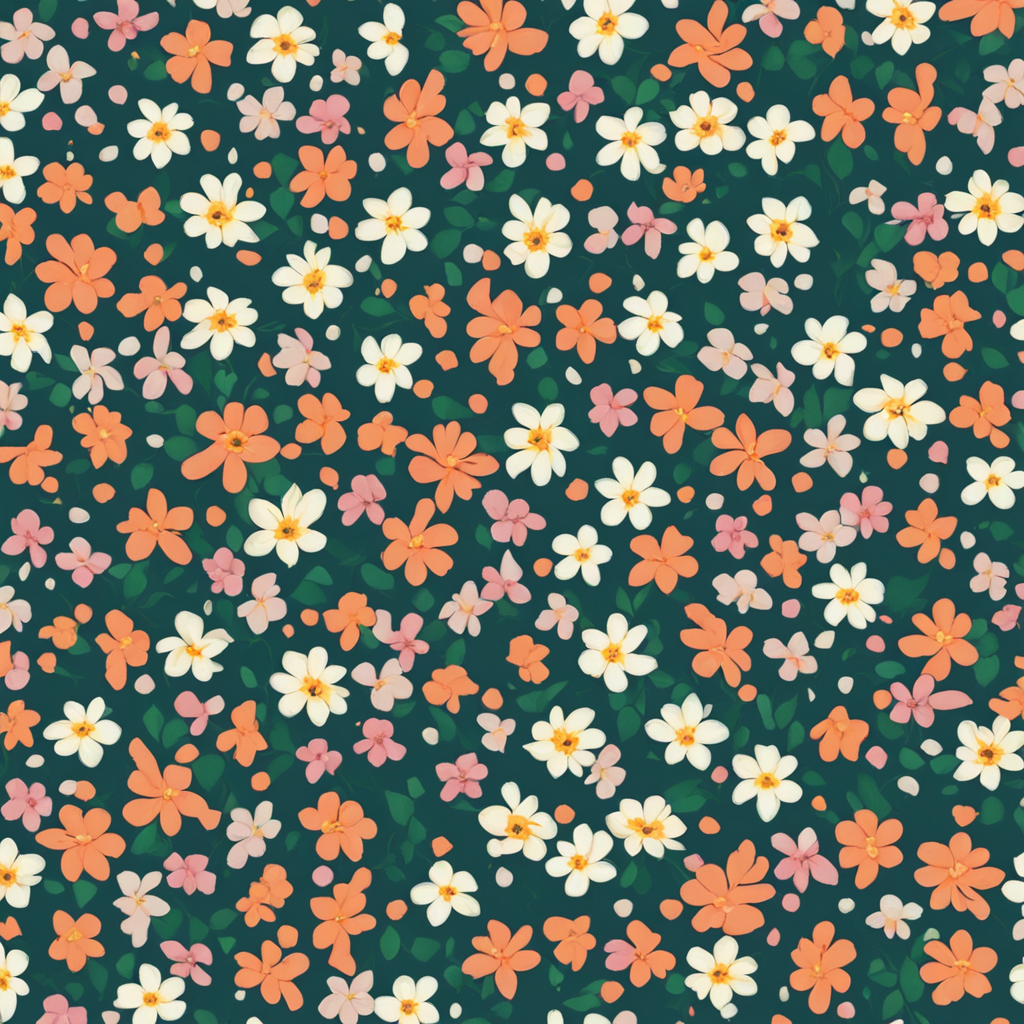}\hfill
    {\Large$=$}\hfill
    \includegraphics[valign=c,width=0.35\textwidth]{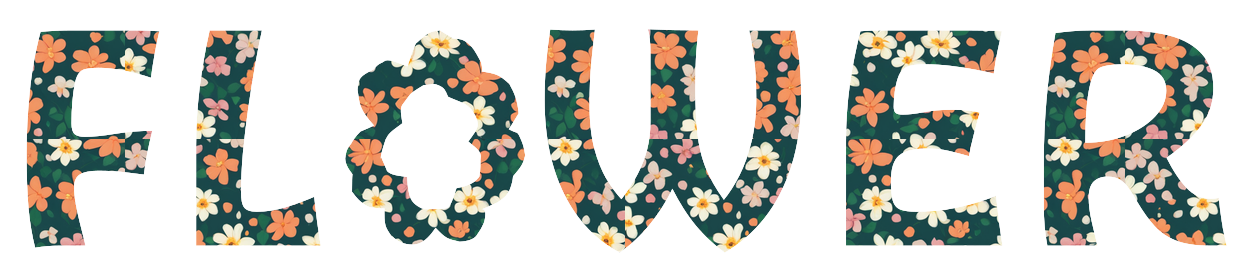}
    \caption{Text Image of Flower + Flower's texture → result.}
    \label{fig:row_a}
  \end{subfigure}
  \begin{subfigure}[b]{\textwidth}
    \centering
    \includegraphics[valign=c,width=0.30\textwidth]{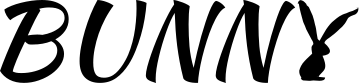}\hfill
    {\Large$+$}\hfill
    \includegraphics[valign=c,width=0.15\textwidth]{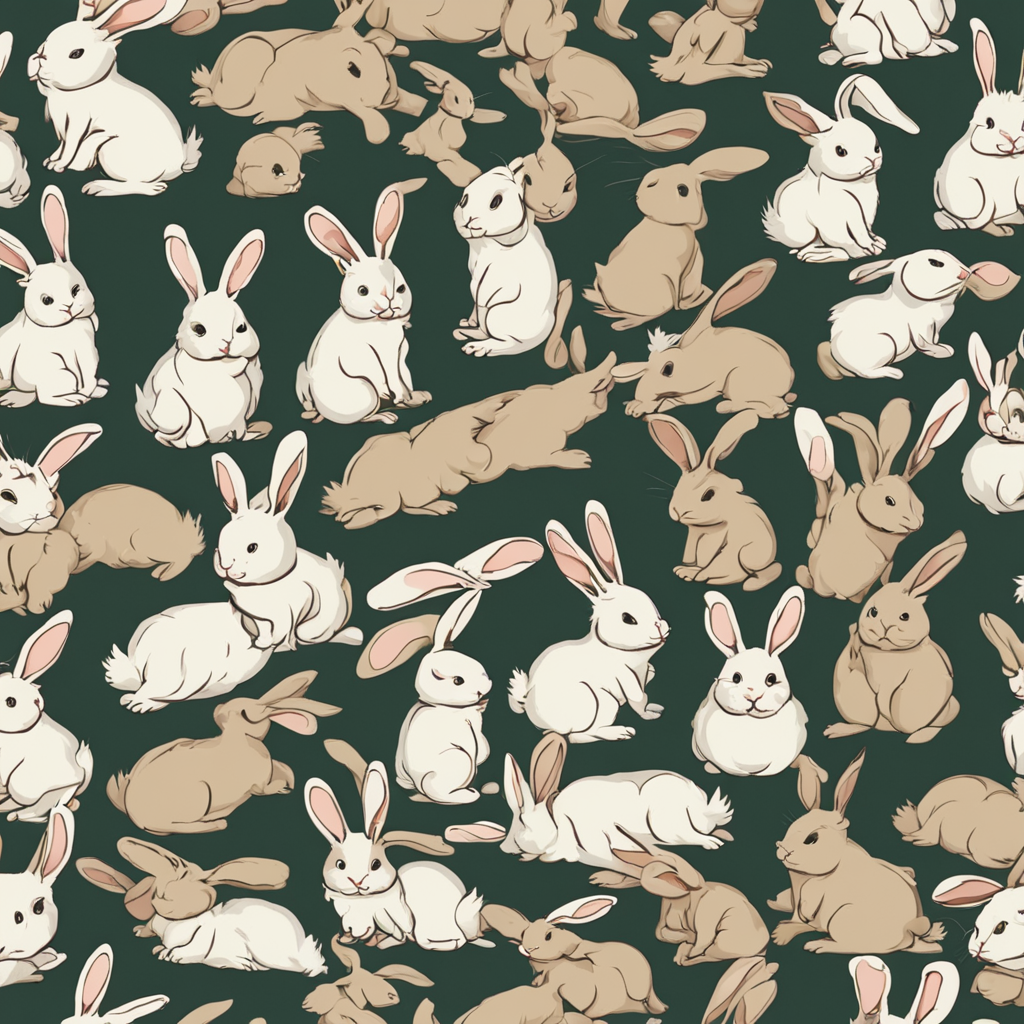}\hfill
    {\Large$=$}\hfill
    \includegraphics[valign=c,width=0.35\textwidth]{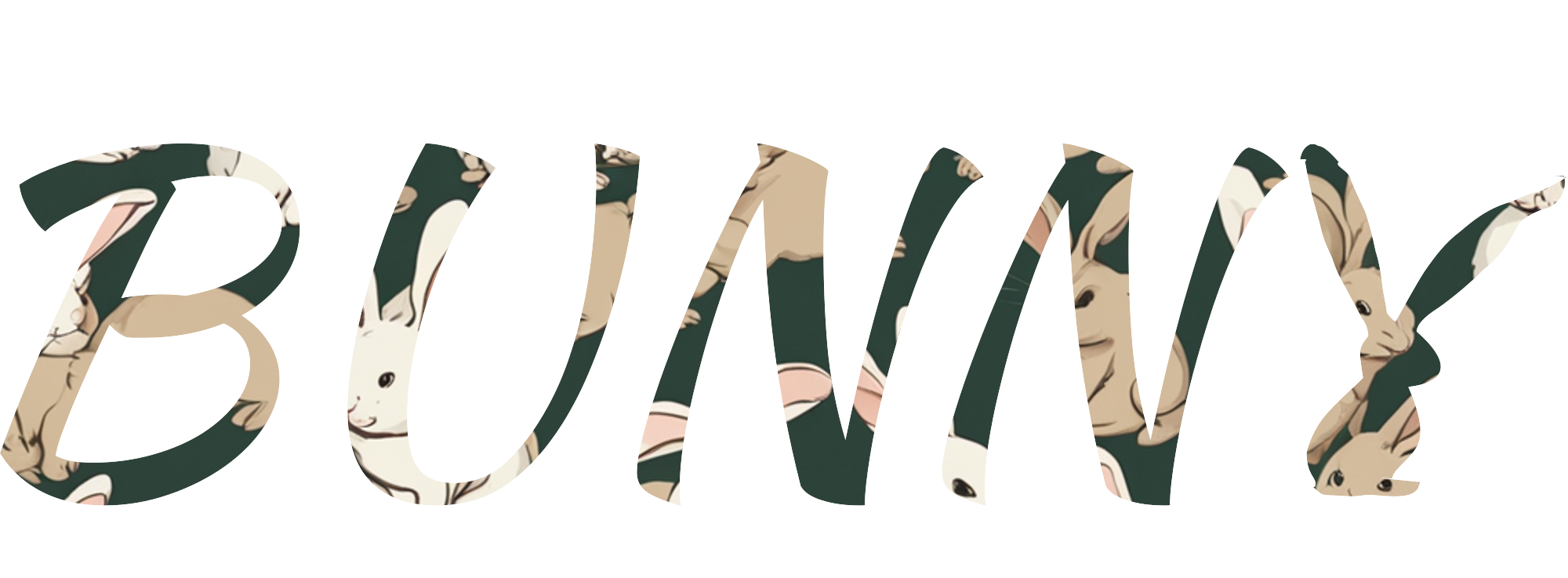}
    \caption{Text image of Bunny + Bunny's texture → result.}
    \label{fig:row_b}
  \end{subfigure}
  \caption{Final outputs of the proposed pipeline based on semantic
concepts, target words, and stylized letters using ”Word as Illustration” mode \cite{iluz2023word}. (a) Uses ”FLOWER” as
both the semantic concept and the word, with the letter ”O” stylized via .
(b) Uses ”BUNNY” as the semantic concept and as the word, with
the letter ”Y” stylized.}
  \label{fig:final_results}
\end{figure*}

Instead of Stable Diffusion altering the letter's structure directly, the model uses Score Distillation Sampling (SDS) \cite{alldieck2024score} to retrieve directions from Stable Diffusion and iteratively refine the letter's control points. SDS adds controlled noise to the rasterized letter, places it in Stable Diffusion, and uses the denoised prediction to compute gradients. These gradients are then used to revise the letter's vector representation, gradually transforming its shape over time to reflect the word's semantic interpretation. Moreover, the authors introduced As-Conformal-As-Possible (ACAP) loss, which prevents excessive deformations and ensures the letter remains legible throughout the transformation process.

\subsection{Texture Generation}
After generating text images, the pipeline generates a subtle texture and pattern using Stable Diffusion XL (SDXL) \cite{podell2023sdxl}. To generate images using Stable Diffusion, our method uses only a positive prompt, which follows the template: "Seamless repeating pattern of tiny and small [semantic concept]"—where the placeholder [semantic concept] is replaced with the user's input from the first text box. Moreover, since textures play a crucial role in this project, and a pattern or subtle texture is required to map onto the text images, a refiner is incorporated into the pipeline to enhance Stable Diffusion’s outputs. This refiner increases detail, which is beneficial for generating high-quality textures.
\subsection{Texture Mapping}
After texture generation, the texture mapping process begins by converting the SVG representation of the text image into a high-resolution PNG image to preserving transparency and to use the alpha channel of the image as a mask in the future steps. The mask $M(x, y)$ is defined as:

\begin{equation}
    M(x, y) = \begin{cases} 
    1, & \text{if pixel } (x, y) \text{ belongs to the text shape} \\ 
    0, & \text{otherwise}
    \end{cases}
\end{equation} 

Then, the generated texture is mapped to the text image as follows. At first, the texture is resized to a smaller scale $s$, creating new texture $T'(x, y)$ and tiled across a blank canvas matching the dimensions of the text image to ensure seamless coverage. Mathematically this can be represented as:
\begin{equation}
    T'(x, y) = T\left(\frac{x}{s}, \frac{y}{s}\right).
\end{equation}

The resizing or scaling algorithm employs the LANCZOS downsampling method, a high-quality resampling filter known for minimizing aliasing and preserving details during scaling \cite{duchon1979lanczos}. Subsequently, the alpha channel of the text image, which represents its transparency, is extracted to create a grayscale mask. This mask serves as a guide for applying the texture, ensuring that the texture fills only the visible regions of the text while maintaining transparency in the background. The textured text is then composited onto a user-defined background color $B$, which is white by default, creating a final image $I_{text}(x, y)$ where the texture $T'(x, y)$ seamlessly integrates with the text while retaining its original shape and visual clarity. The final output generation is formulated using the following equation:
\begin{equation}
    I_{text}(x, y) = M(x, y) \cdot T'(x, y) + (1 - M(x, y)) \cdot B.
\end{equation}

This method effectively maps the texture onto the text, producing an output that is both aesthetically pleasing and highly readable. Examples of the final result of our proposed pipeline are shown in \cref{fig:final_results}, where the generated texture is mapped to original generated text images to produce the final stylized artistic text styles.

\subsection{User Interaction}
As mentioned earlier, after generating the final output, if the user is unsatisfied with the texture's scale or background color, they can make edits using an editorial panel designed for this purpose. This panel is displayed to the user upon clicking "Adjust Image." It includes a color picker that allows the user to set a new hex color for the background of the generated image. Additionally, the editorial panel features a slider for adjusting the texture's scale. The user can scale down the texture to better map onto the text. The user interface has been developed using the Gradio framework.

\section{Results}
To generate the results, we used an NVIDIA A100 GPU with 40 GB of VRAM. The project employs Stable Diffusion XL for texture generation, while the primary part of the pipeline, based on the work of Iluz et al. (2023) \cite{iluz2023word}, utilizes the Stable Diffusion model \cite{rombach2022high} to modify the shapes of the letters. In this project, letter shape generation was configured to run for 1,500 iterations, while texture generation using Stable Diffusion XL was set to 500 iterations. Under these settings, the generation of each image required approximately 9 minutes and 40 seconds to complete.

We used a structured questionnaire to assess the results of our method. We invited 18 participants to answer the questions and evaluate the generated outputs. Each participant observed eight images generated by the pipeline, corresponding to the semantic concepts: "BOOK," "CHAIR," "FISH," "HOUSE," "SNAKE," "SNOW," "SUNFLOWER," and "SURFING." They were asked to evaluate each image based on two key questions:
\begin{itemize}
    \item \textbf{How well does the texture represent the meaning of the word?}
    \item \textbf{Does the texture distract from the readability of the word?}
\end{itemize}
The first question addresses the quality of the semantic component, while the second focuses on the legibility of the generated outputs. The answer set for the first question includes "Not at all," "Slightly," "Moderately," "Very well," and "Extremely well." The options for the second question consist of "No," "Somewhat," and "Yes."

\begin{figure}[ht]
    \centering
    \includegraphics[width=0.8\linewidth]{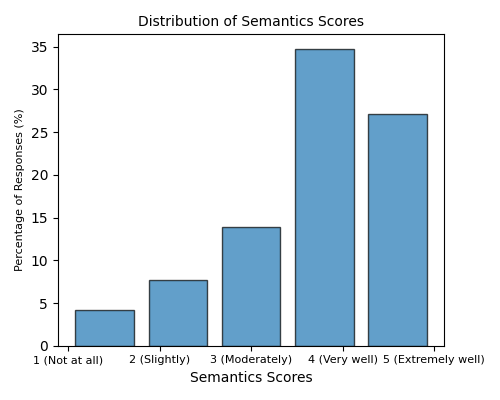}\\[1ex]
    \caption{Distribution ratings for how well the generated texture conveys the semantic concepts of the final generated outputs. As it is show, roughly 65\% of the testers have indicated strong alignment between the textures and semantic concepts.}
    \label{fig:statistics}
\end{figure}

The feedback revealed a clear trend in how well the textures conveyed semantic meanings, while having no significant effect on the legibility of the text images generated using the "Word-As-Image" model. The distribution of user ratings regarding the functionality and effectiveness of the proposed pipeline is shown in \cref{fig:statistics}.

\cref{tab:results} presents the evaluation results of the generated output based on users' responses. The semantic column indicates how well users felt the texture represented the meaning of the semantic concept, whereas the legibility column shows the extent to which the texture affected the readability of the text images. The reported values are based on means and standard deviations, representing average user ratings and variability in responses, respectively. Additionally, we provided the "Word-As-Image" results alongside our statistics to demonstrate that adding texture to their outputs has had little to no effect on the legibility of the text images.

\begin{table}[ht]
\centering
\caption{Evaluation results of generated output based on the semantics and legibility. Results of "Word-As-Image" model is written to show texture mapping has a few negative impact on the legibility}
\begin{tabular}{lcc}
\toprule
\textbf{Method} & \textbf{Semantics} & \textbf{Legibility} \\
\midrule
Ours & 0.722 (0.253) & 0.871 (0.256) \\
World-As-Image & 0.8 (0.191) & 0.9 (0.170)\\
\bottomrule
\end{tabular}
\label{tab:results}
\end{table}

\begin{figure}[htb]
    \centering
     \includegraphics[width=1\linewidth]{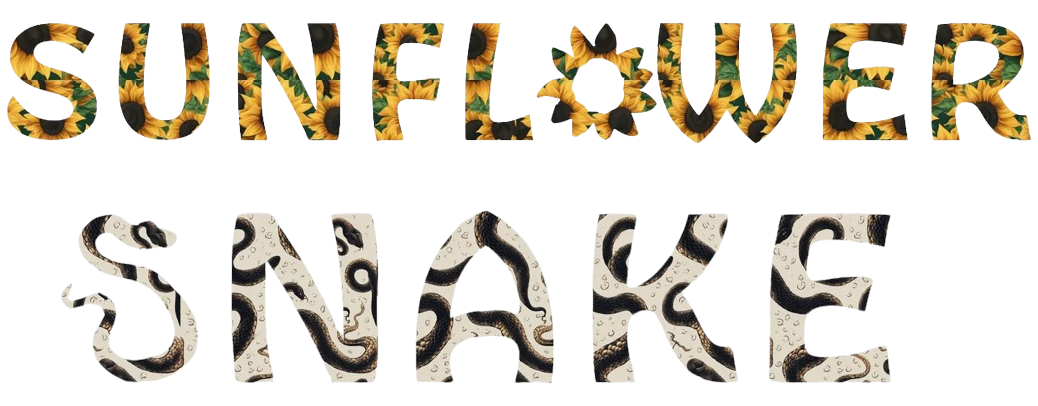}
    \caption{
           66.7 percent of the participants have stated that texture of these two images better represented the meaning of semantic concept. }
    \label{fig:best_results}
\end{figure}

Among all concepts, the textures for "SUNFLOWER" and "SNAKE" received the highest ratings, with 44.4\% of participants indicating that the textures represented the concepts "Extremely well." These results are illustrated in \cref{fig:best_results}, showing the texture for "SUNFLOWER" in the letter "O" and for "SNAKE" in the letter "S."

Overall, participant responses indicate that the pipeline successfully integrates textures into text while maintaining both semantic clarity and readability. The readability scores are shown in \cref{tab:legibility_scores}. This table shows the distribution ratings for how much the textures reduced the readability of the generated text images. Roughly 70\% of the testers have stated no legibility reduction in the final outputs. The results underscore the potential of this approach for creative typography, particularly in applications where a meaningful visual representation of text is essential.

\begin{table}[ht]
  \centering
  \caption{Distribution of Legibility Score. Please note that ”No” means no distraction}
  \label{tab:legibility_scores}
  \begin{tabular}{lrr}
    \toprule
    \bfseries Legibility Score & \bfseries Percentage (\%) \\
    \midrule
    (No)       & 71.53 \\
    (Somewhat) & 11.11  \\
    (Yes)      &  4.86  \\
    \bottomrule
  \end{tabular}
  \label{tab:ligibility scores}
\end{table}


\section{Conclusion and Future works}
\label{sec:5}
In this paper, an automatic pipeline is proposed that takes three inputs— a semantic concept, a word, and a letter— from the user to generate a text image in which the shape of the chosen letter is modified, and a texture based on the entered semantic concept is mapped onto the text. To modify the shape of the letter, the work of Iluz et al. \cite{iluz2023word} was used. Then, to generate the textures, the Stable Diffusion model was utilized. The output of this pipeline can be useful in various design workflows, such as graphic design, poster design, and banner design \cite{tatsukawa2024fontclip,bai2024intelligent}.

To analyze the generated images and evaluate the pipeline, a questionnaire was designed and administered to 18 participants. The proposed pipeline has some limitations inherited from the models it uses. This method relies on a pretrained Stable Diffusion model for texture generation, while \cite{iluz2023word} also used a pretrained Stable Diffusion model to modify the shape of the letters. While pretrained models are highly versatile within the domains they have been trained on, they encounter challenges when applied to new domains. As a result, this pipeline has limited knowledge and may struggle to generate legible and meaningful results for abstract concepts. One possible solution for future work is to retrain or fine-tune the mentioned models on a broader dataset encompassing a wider range of semantic concepts.

\section{SUPPLEMENTAL MATERIALS}
\label{sec:6}
All supplemental materials are available on \url{https://doi.org/10.5281/zenodo.15275234}. Specifically, the supplemental materials include: a PDF version of the questionnaire, the images used during the tests, the raw user response data, and a demonstration video of our method.

\bibliographystyle{abbrv-doi}
\bibliography{template}
\end{document}

%% file: figures/dataflowdiagram.tex
\usetikzlibrary{shapes.geometric, arrows, positioning, calc, fit, backgrounds}

\tikzstyle{input} = [rectangle, rounded corners, minimum width=2.5cm, minimum height=1cm, draw=black, fill=red!40, text centered, align=center]
\tikzstyle{module} = [rectangle, minimum width=4cm, minimum height=1.1cm, text centered, draw=black, fill=purple!30]
\tikzstyle{developed_module} = [rectangle, minimum width=4cm, minimum height=1.1cm, text centered, draw=black, fill=red!40]
\tikzstyle{group} = [draw=black, dashed, inner sep=0.5cm]
\tikzstyle{arrow} = [thick,->,>=stealth]
\tikzstyle{output} = [rectangle, rounded corners, minimum width=3cm, minimum height=1cm, draw=black, fill=orange!30, text centered, align=center]
\tikzstyle{dashedarrow} = [->, dashed, thick, >=stealth]

\begin{tikzpicture}[node distance=2cm, auto]

\node (concept) [input] {Semantic Concept \\ \textbf{"TREE"}};
\node (word) [input, below of=concept] {Word \\ \textbf{"NATURE"}};
\node (letter) [input, below of=word] {Letter \\ \textbf{"T"}};

\begin{scope}[on background layer]
\node[group, fit=(concept) (word) (letter)] (userinputs) {};
\end{scope}
\node[above of=userinputs, yshift=1.5cm, font=\bfseries] {User Inputs};

\node (stable_diffusion) [module, right of=concept, xshift=4cm] {Stable Diffusion XL Module};

\node (img_stable) [above of=stable_diffusion, node distance=2cm] 
    {\includegraphics[width=4cm]{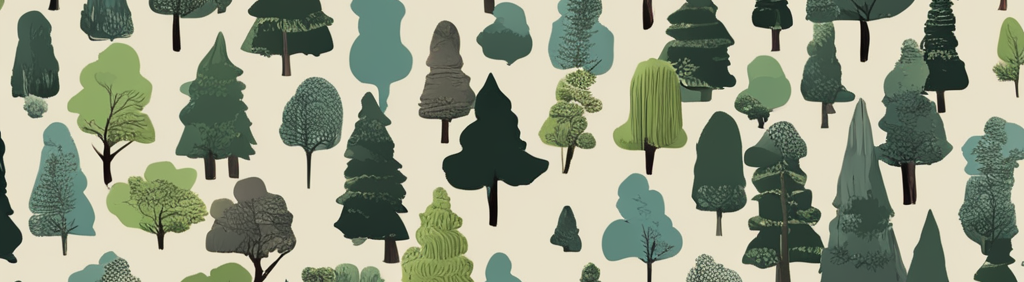}};
\draw [dashedarrow] (stable_diffusion.north) -- (img_stable.south);

\node (word_as_image) [module, below of=stable_diffusion, yshift=-1.25cm] {Word-As-Image Module};

\node (img_word) [below of=word_as_image, node distance=1.5cm]
    {\includegraphics[width=3cm]{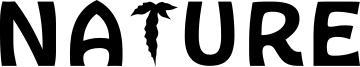}};
\draw [dashedarrow] (word_as_image.south) -- (img_word.north);

\node (texture_mapping) [developed_module, right of=stable_diffusion, xshift=4cm]{Texture Mapping};

\node (texture_edit) [developed_module, below of=texture_mapping, yshift=0.15cm, align=center]{Texture scaling and \\ background color change};

\node (final_output) [output, right of=word_as_image, xshift=4cm, yshift=-10]{Final Output};

\node (img_final) [below of=final_output, node distance=1.25cm]
    {\includegraphics[width=3.5cm]{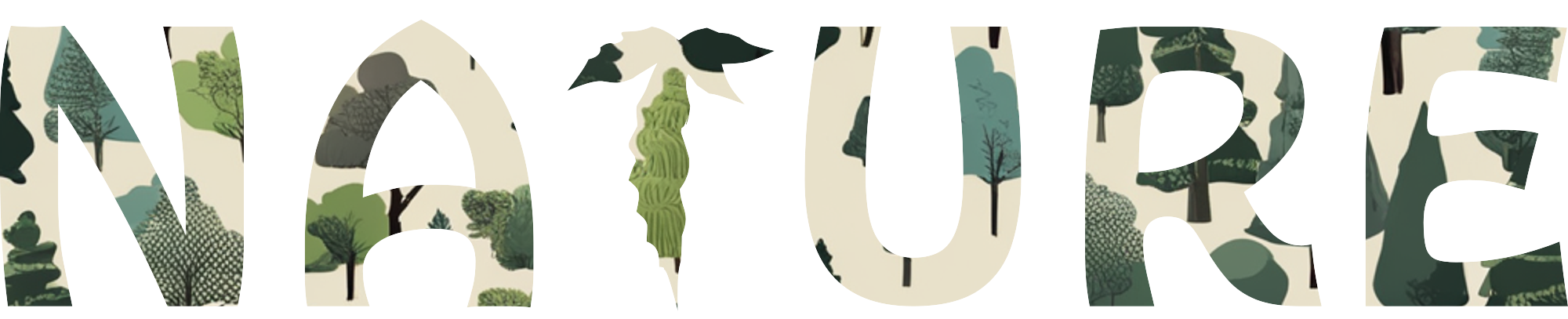}};

\draw [dashedarrow] (final_output.south) -- (img_final.north);

\draw [arrow] (concept.east) -- ++(1,0) |- (stable_diffusion.west);
\draw [arrow] (concept.east) -- ++(1,0) |- (word_as_image.west);

\draw [arrow] (word.east) -- ++(1,0) |- (word_as_image.west);
\draw [arrow] (letter.east) -- ++(1,0) |- (word_as_image.west);

\draw [arrow] (word_as_image.east) -- ++(1.5,0) |- (texture_mapping.west); 
\draw [arrow] (stable_diffusion.east) -- ++(1,0) |- (texture_mapping.west);

\draw [arrow] (texture_mapping.south) -- (texture_edit.north);
\draw [arrow] (texture_edit.south) -- (final_output.north);

\node[below of=stable_diffusion, yshift=1.15cm, font=\normalsize\itshape] {Generates Subtle Texture};
\node[above of=word_as_image, yshift=-1.15cm, font=\normalsize\itshape] {Generates Text Image with Stylized Letter};

\end{tikzpicture}